# Magnetic ordering in pressure-induced phases with giant spin-driven ferroelectricity in multiferroic TbMnO$_3$

Noriki Terada,[1,*] Dmitry D. Khalyavin,[2] Pascal Manuel,[2] Toyotaka Osakabe,[3] Akiko Kikkawa,[1] and Hideaki Kitazawa[1]

[1]*National Institute for Materials Science, Sengen 1-2-1, Tsukuba, Ibaraki 305-0047, Japan*
[2]*ISIS Facility, STFC Rutherford Appleton Laboratory, Chilton, Didcot, Oxfordshire, OX11 0QX, United Kingdom*
[3]*Japan Atomic Energy Agency, Tokai, Ibaraki 319-1195, Japan*

In order to clarify the mechanism associated with pressure/magnetic-field-induced giant ferroelectric polarization in TbMnO$_3$, this work investigated changes in magnetic ordering brought about by variations in temperature, magnetic field, and pressure. This was accomplished by means of neutron diffraction analyses under high pressures and high magnetic fields, employing a single crystal. The incommensurate magnetic ordering of a cycloid structure was found to be stable below the reported critical pressure of 4.5 GPa. In contrast, a commensurate $E$-type spin ordering of Mn spins and a noncollinear configuration of Tb spins with $\bm{k} = (0,\frac{1}{2},0)$ appeared above 4.5 GPa. The application of a magnetic field along the $a$ axis ($H_{\|a}$) under pressure induces a $\bm{k} = (0,0,0)$ antiferromagnetic structure in the case of Tb spins above $H_{\|a}$, enhancing the ferroelectric polarization, while the $E$-type ordering of Mn spins is stable even above the critical field. From the present experimental findings, we conclude that the $E$-type ordering of Mn spins induces giant ferroelectric polarization through an exchange striction mechanism. The $H_{\|a}$-induced polarization enhancement can be understood by considering that the polarization, reduced by the polar ordering of Tb moments in a zero field, can be recovered through a field-induced change to nonpolar $\bm{k} = (0,0,0)$ ordering at $H_{\|a} \sim 2$ T.

Spin orderings/correlations in magnetic materials can be controlled by the application of pressure, since changing the bond distances and angles generally affects exchange interactions. When one magnetic order parameter is coupled to another, it is thus possible to vary other physical quantities through pressure-tuned spin orderings. In multiferroic materials, characterized by strong coupling between (anti)ferromagnetism and ferroelectricity [1–3], the application of external perturbations such as magnetic fields and pressure can significantly affect the ferroelectric properties by modifying the magnetic ordering or symmetry [4–12]. Recently, Aoyama *et al.* reported the pressure-induced giant ferroelectric polarization of the orthorhombic rare-earth perovskite TbMnO$_3$ [10]. This work identified a flop of the polarization along the $c$ axis, $P_{\|c}$, to the $a$ axis at the highest value $P_{\|a} \sim 1.0$ $\mu$C/cm$^2$ reported for spin-driven multiferroics, through the application of pressures above 4.5 GPa. The schematic phase diagram for this material is shown in Fig. 1(a) [10]. The application of a magnetic field along the $a$ axis, $H_{\|a}$, was also found to enhance the polarization above $H_{\|a} \sim 2$ T, up to a maximum of $P_{\|a} \sim 1.8$ $\mu$C/cm$^2$. However, the variations in the ferroelectric polarization of TbMnO$_3$ with pressure and magnetic field are not yet fully understood, due to a lack of magnetic structural determination and a scarcity of information regarding magnetic characteristics under various magnetic fields and pressures.

In previous powder neutron diffraction studies of TbMnO$_3$ under pressure, $E$-type ordering was observed above 0.9 GPa [5], which is inconsistent with polarization measurements [10]. In the present work, in order to clarify the mechanism associated with the giant ferroelectric polarization flop and enhancement phenomena, we investigated the variation of magnetic orderings induced in TbMnO$_3$ by changing the temperature, magnetic field, and pressure, by means of neutron diffraction experiments under high pressures and high magnetic fields, employing a single crystal.

In preparation for these studies, a single crystal sample, grown by the floating-zone method, was cut into a platelike shape with dimensions $0.6 \times 0.5 \times 0.2$ mm$^3$. The neutron diffraction work was performed using a recently developed apparatus consisting of a combination of a hybrid anvil-type high pressure cell [13,14] and a cold neutron (time of flight) diffractometer [15] at the ISIS Facility in the UK. We used a 4:1 mixture of methanol-ethanol and glycerin as pressure media for the experiments for 2.1, 4.2, and 5.0 GPa, respectively. Pressure values were determined by the ruby fluorescence method at room temperature. The pressure cell was inserted into the vertical field superconducting magnet such that the $a$ axis ($Pbnm$ setting) was vertical. Magnetic structure refinements were performed using the FULLPROF program [16].

Typical TbMnO$_3$ magnetic neutron diffraction profiles obtained under pressures up to 5.0 GPa and at low temperature are shown in Fig. 2. Magnetic propagation vectors at ambient pressure have been reported in previous neutron diffraction work to be incommensurate below $T_N^{Tb} = 7$ K: $\bm{k}_{Mn} = (0,0.28,0)$ and $\bm{k}_{Tb} = (0,0.42,0)$ [17,18]. With increasing pressure, $\bm{k}_{Mn}$ changes to $(0,0.31,0)$ at 2.1 GPa and $(0,0.35,0)$ at 4.2 GPa, while $\bm{k}_{Tb}$ remains unchanged at $(0,0.42,0)$ below the phase transition pressure of 4.5 GPa [10]. These results suggest that the cycloid spin structure at ambient pressure is stable up to at least 4.2 GPa. In contrast, the peak position is drastically changed at 5.0 GPa, to $(0,1.5,0)$, indicating that the propagation vectors for both the Mn and Tb spins both lock into a value of $\bm{k} = (0,\frac{1}{2},0)$ above 4.5 GPa. Additional reflections were also observed at $(0,\frac{1}{2} \pm m,n)$ (where $m$ and $n$ are integers) at 5.0 GPa and 1.5 K, as seen in the contour intensity map shown in Fig. 3(a).

*TERADA.Noriki@nims.go.jp

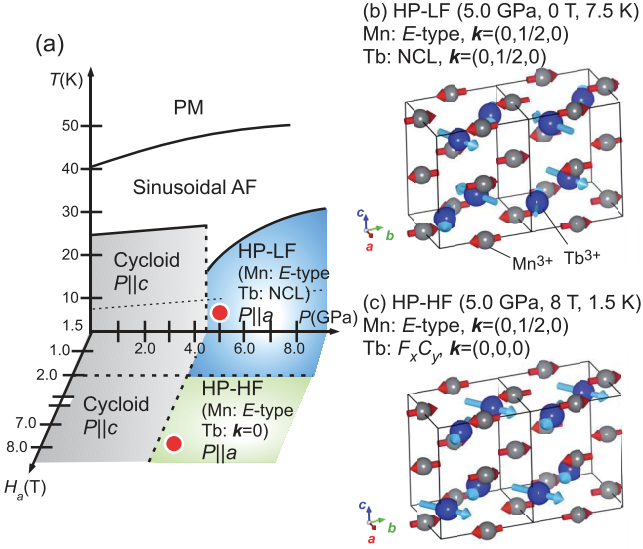

FIG. 1. (a) Schematic illustration of the magnetic phase diagram of TbMnO$_3$ [10]. The magnetic structures determined for (a) the high pressure, low field (HP-LF) and (b) high pressure, high field (HP-HF) phases.

It is evident from Fig. 4(a) that $(0, \frac{1}{2} \pm m, n)$ reflections for which $n = $ odd and $n = $ even were observed below $T \sim 10$ K and $H = 0$ T, corresponding to the temperature at which the Tb moments become ordered. The line shape of the diffraction patterns at $H = 0$ T and $P = 5.0$ GPa below $T \sim 10$ K depend on the cooling process: a three-peak structure around the commensurate position is observed after zero field cooling (ZFC), while a single peak is generated at the commensurate position after magnetic field cooling (FC), as shown in Fig. 5. The different line shapes obtained from the ZFC and FC processes have not yet been fully explained. They may originate from the coexistence of a low pressure incommensurate phase caused by incomplete phase transition under ZFC conditions. Above $T \sim 10$ K, only $(0, \frac{1}{2} \pm m, n)$ with $n = $ odd reflections is present, indicating that the magnetic ordering of the Mn moments involves antiferromagnetic stacking along the $c$ axis, which is consistent with the $E$-type structure in orthorhombic HoMnO$_3$ [20,21].

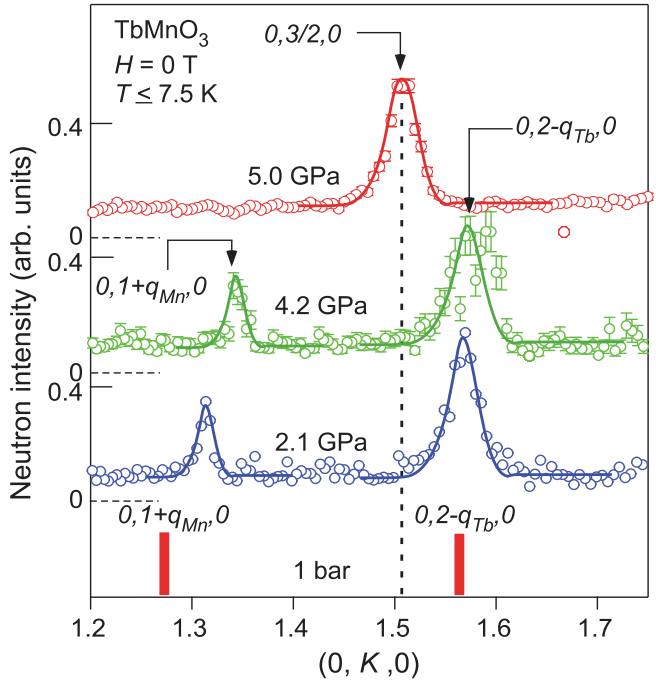

FIG. 2. Pressure dependence of the neutron diffraction pattern along the $(0, K, 0)$ reciprocal lattice line. The 2.1 and 4.2 GPa data were acquired after cooling at zero magnetic field, while the data at 5.0 GPa were acquired at 7.5 K after cooling at 8 T. The vertical bars in these plots denote the positions expected at ambient pressure.

With increasing $H_{\|a}$ at $T = 1.5$ K and 5.0 GPa, the diffraction patterns remain almost unchanged up to 1 T, while $H_{\|a}$ changes drastically above 2 T. The critical magnetic field is coincident with the field at which the ferroelectric polarization is enhanced [10]. Taking the $H_{\|a}$ response of TbMnO$_3$ at ambient pressure into consideration [22], it is evident that the significant changes in neutron intensities at $H_{\|a} = 2$ T are associated with a metamagnetic phase transition of the Tb spins. Some additional reflections having integer indexes and forbidden in the $Pbnm$ space group (e.g., 0,1,1 and 0,1,2) were observed above $H_{\|a} = 2$ T, as shown in Figs. 3(b), 4(c), and 5. This corresponds to the onset of antiferromagnetic ordering, with $\bm{k} = (0,0,0)$ induced by the magnetic field. The space-group-allowed reflection, $0, -2, 1$, was also significantly enhanced at the critical magnetic field, indicating a field-induced ferromagnetic component along the $a$ axis. Conversely, as can be seen from Figs. 4(c) and 5, the commensurate peak intensities at $(0, \frac{1}{2} \pm m, n)$ ($n = $ even) disappear and $n = $ odd reflections are significantly reduced at $H_{\|a} = 2$ T. These results indicate that the Tb spins order into a $\bm{k} = (0,0,0)$ structure above 2 T, while the Mn moments retain $\bm{k} = (0, \frac{1}{2}, 0)$ ordering even above the critical field.

As shown in Fig. 4(b), $(0, \frac{1}{2} \pm m, n)$ ($n = $ odd) reflections were observed only below $T \sim 30$ K at $H = 8$ T and $P = 5.0$ GPa, corresponding to the reported ferroelectric phase transition point [10,11]. In the case of the $\bm{k} = (0,0,0)$ reflections, the 0,1,1 forbidden reflection disappeared at 60 K, while the $(0, -2, 1)$-allowed reflection, corresponding to the ferromagnetic component, remained even at the highest experimental temperature (60 K). At $T = 30$ K, $H = 8$ T, and 5.0 GPa, a reflection was observed at $(0, -0.4, 3)$ in addition to the $\bm{k} = (0,0,0)$ reflections, possibly corresponding to incommensurate spin-density-wave Mn spin ordering [see Fig. 4(b) in the Supplemental Material].

In order to determine the magnetic structures in pressure-induced phases, we performed a magnetic structure analysis using the magnetic (and nuclear) reflections generated at 5.0 GPa [23]. This analysis identified two two-dimensional irreducible representations (IRs), $mX_1$ and $mX_2$, for $\bm{k} = (0, \frac{1}{2}, 0)$ at the Mn(4$a$) and Tb(4$c$) sites in the high pressure, low field (HP-LF) phase (Ref. [24]). During this analysis, we explored different symmetry distinct combinations of the order parameter directions (OPDs) in the reducible $mX1 \oplus mX2$ representation space, and selected those that restrict the electric polarization such that it is along the $a$ axis. The optimum refinement was obtained using $mX_1$ for Mn spins and $mX_2$ for Tb spins. In the case of the Mn moments, one of modes for an OPD, $P3(a,a)$, in $mX_1$ corresponds to a ↑↑↓↓ spin configuration of $b$ spin components, representing a so-called $E$-type antiferromagnetic structure with a Mn moment of $4.0\mu_B$. The magnetic space group for the $E$-type ordering was polar $P_bmn2_1$ with a basis $\{(0,0,1),(0,1,0),(1,0,0)\}$ of the original $Pbnm$ setting. For Tb spins, selecting one of the OPDs, $P3(a,a)$, for the $mX_2$ IR gave the best refinement with a combination of the $E$-type ordering of Mn spins. As illustrated in Fig. 1(b), the magnetic ordering determined for Tb spins was a noncollinear structure with spin components $M = [5.9(3),1.8(8),0]\mu_B$ in the $ab$ plane. The quality of refinement for the HP-LF phase ($R_F = 7.1\%$) is presented in Fig. 3(c). The determined Mn and Tb orderings in the HP-LF phase can be expressed as a combination of two IRs, $mX_1 \oplus mX_2$ [for which the OPD is $P3(1)P3(1)(a,a;b,b)$], with the magnetic space group $P_a2_1$ (allowing polarization only along the $a$ axis in a $Pbnm$ setting).

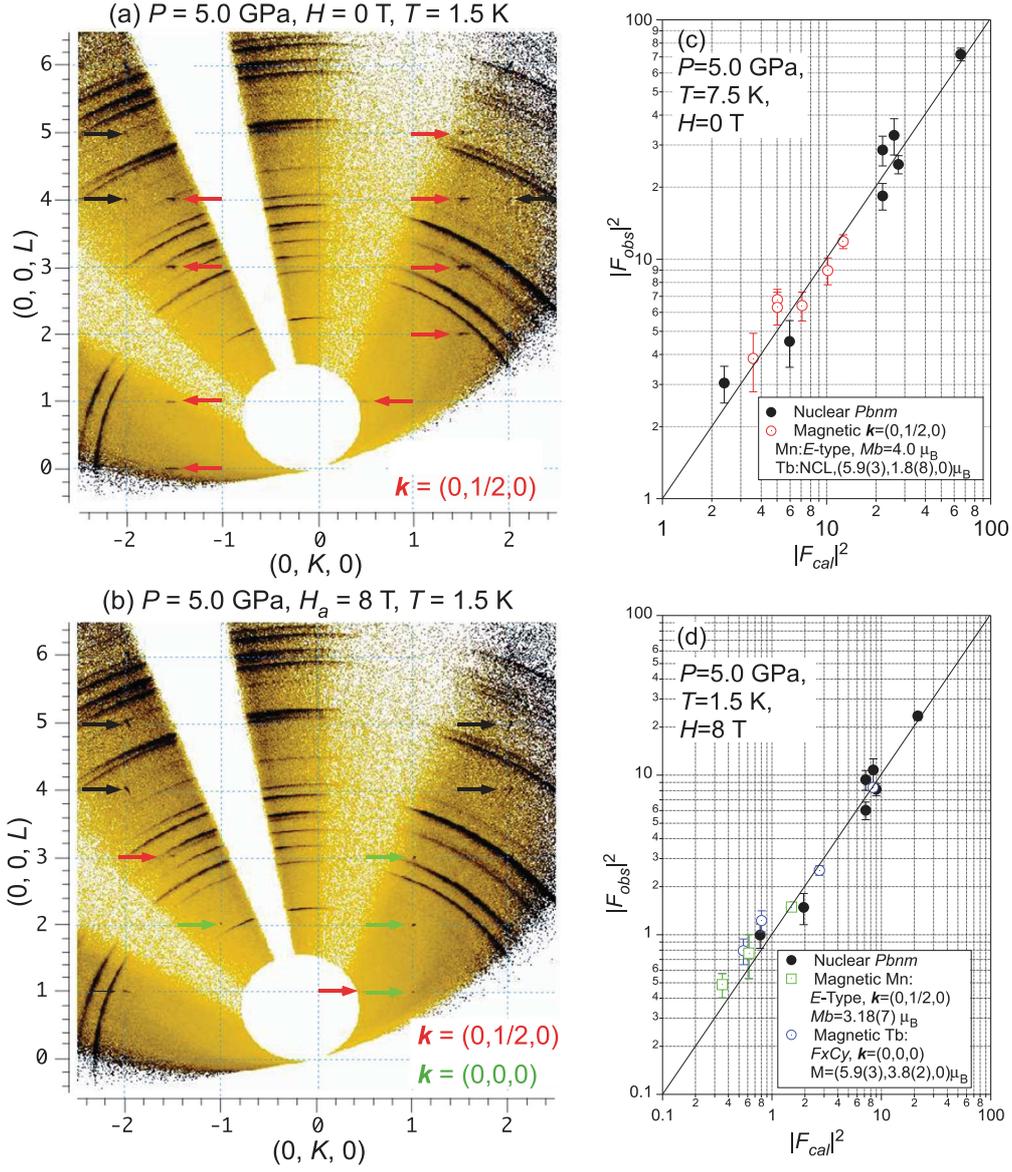

FIG. 3. Contour maps of neutron intensity in the $(0,K,L)$ reciprocal zone at 5.0 GPa, 1.5 K and (a) 0 T and (b) 8 T. Arrows indicate positions of observed nuclear and magnetic Bragg reflections. Results of the magnetic structure refinements for the data at (c) $T = 7.5$ K, $H = 0$ T, and 5.0 GPa, and (d) $T = 1.5$ K, $H = 8$ T, and 5.0 GPa. The experimental data were corrected for absorption and Lorentz factor. The reliability factors for the magnetic parts, $R_F = 7.1\%$ at $H = 0$ T and $R_F = 6.5\%$ at $H = 8$ T, were obtained for the model shown in Figs. 1(b) and 1(d), respectively. The insets show the refined parameters.

In the high pressure and high magnetic field phase (HP-HF), magnetic reflections were observed for $(0,\frac{1}{2} \pm m,n)$ ($n =$ odd) in the case of Mn ordering and at $Pbnm$ allowed and forbidden integer indexed positions for Tb spins. The refinement of the $E$-type Mn ordering was in good agreement with the HP-HF phase, as shown in Fig. 3(d). In the case of the Tb moments, there were eight one-dimensional IRs for $k = (0,0,0)$ associated with the little group of $Pbnm$. Only one of the IRs, $m\Gamma_2^+$ ($F_x,C_y$ in Bertaut's notation [25]), allows a ferromagnetic component along the $a$ direction, and corresponds to the nonpolar $Pbn'm'$ magnetic space group. When we combine $m\Gamma_2^+$ for Tb spins with $mX_1$ for Mn spins, to obtain $m\Gamma_2^+ \oplus mX_1$, the resulting magnetic space group is $Pn'a2_1$ (with basis $\{(0,2,0),(0,0,1),(1,0,0)\}$ in a $Pbnm$ setting), allowing ferroelectric polarization only along the $a$ axis in $Pbnm$. The calculated structure factors exhibit a good fit with experimental data with $R_F = 6.5\%$, as shown in Fig. 3(d). The Tb and Mn spin configurations are illustrated in Fig. 1(c). The magnetic momentum components were refined to be $M_b(\text{Mn}) = 3.18(7)\mu_B$ (assuming a zero ferromagnetic component for Mn spins) and $\boldsymbol{M}(\text{Tb}) = [5.9(3), 3.8(2), 0]\mu_B$.

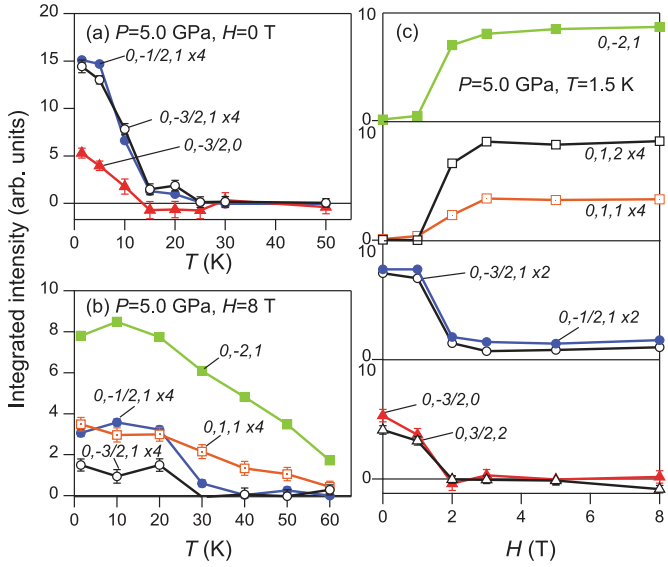

FIG. 4. Temperature dependence of the integrated intensity of observed reflections at (a) $H = 0$ T and (b) $H = 8$ T at 5.0 GPa. (c) Magnetic field variation of the integrated intensity at $T = 1.5$ K and 5.0 GPa. The raw diffraction patterns before integration are provided in the Supplemental Material [19].

Here it is helpful to discuss the pressure and magnetic field variations of ferroelectric polarization reported by Aoyama et al. [10,11]. Theoretical studies have determined that an exchange striction mechanism can explain the significant ferroelectricity associated with $E$-type ordering [26,27]. Aoyama et al. found that the application of a magnetic field ($H_{\|a}$) enhances the electric polarization from $\sim 0.8$ $\mu$C/cm$^2$ to $\sim 1.3$ $\mu$C/cm$^2$ at a $H_{\|a} \sim 2$ T at 7.1 GPa [10,11]. In addition, our own neutron diffraction data demonstrate that the magnetic order of Tb spins changes from a polar noncollinear structure to a nonpolar $\boldsymbol{k} = (0,0,0)$ antiferromagnetic state. We can thus propose one possible scenario for magnetic field changes in ferroelectric polarization induced by pressure in TbMnO$_3$. We suggest that, in a zero field, the $E$-type Mn order induces polarization below $\sim 25$ K, while the polar Tb ordering reduces the polarization value in the opposite direction to compensate for the ferroelectric distortion and reduce the elastic energy below $\sim 10$ K. The application of a magnetic field changes the magnetic order of Tb spins from polar to nonpolar, and so the ferroelectric polarization can be recovered above $H_{\|a} \sim 2$ T. In fact, the reported polarization value at $H = 0$ T and 7.1 GPa decreases from a maximum of $\sim 1.3$ $\mu$C/cm$^2$ to $\sim 0.8$ $\mu$C/cm$^2$ below $\sim 10$ K, while the maximum value is evidently recovered by the application of a magnetic field [10,11]. To obtain a better understanding of the field variations of polarization under pressure in TbMnO$_3$, detailed theoretical calculations including Tb ordering will be required.

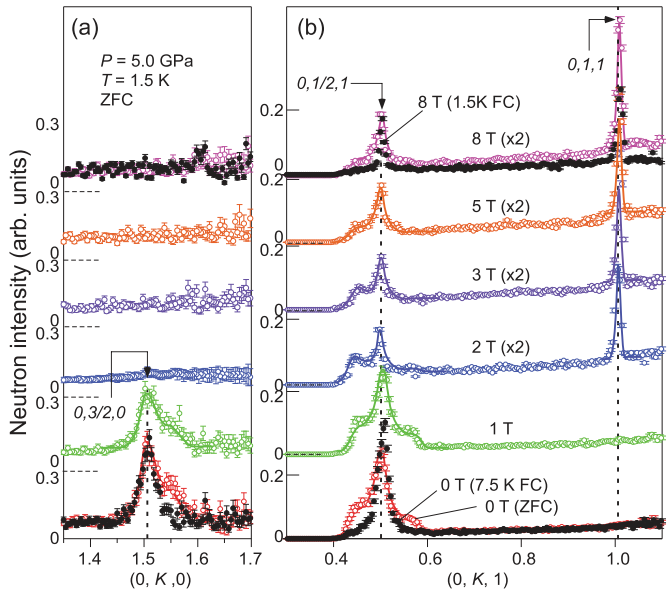

FIG. 5. Magnetic field dependence of the magnetic neutron diffraction pattern along the (a) $(0, K, 0)$ and (b) $(0, K, 1)$ lines at $T = 1.5$ K and 5.0 GPa. Open and closed symbols denote data acquired after ZFC and FC processes, respectively.

In orthorhombic $R$MnO$_3$ ($R = $ Eu-Lu), the magnetic ground states change in the sequence $A$-type (Eu, Gd) to cycloid (Tb, Dy) to $E$-type (Ho, Tm, Yb, Lu) with decreasing ionic radius [28,29]. Thus, when the cell volume in TbMnO$_3$ decreases through substitution at the rare-earth sites, corresponding to the application of chemical pressure, the magnetic ground state of the cycloid order should enter the $E$-type region. Even in this cycloidal phase, the magnetic propagation vector $\boldsymbol{k}$ changes from $(0, 0.28, 0)$ in TbMnO$_3$ to $(0, 0.41, 0)$ in DyMnO$_3$ in response to chemical pressure. Physical pressure also changes $\boldsymbol{k}$, to $(0, 0.31, 0)$ at 2.1 GPa and $(0, 0.35, 0)$ at 4.2 GPa, and induces $E$-type antiferromagnetic ordering above 4.5 GPa. These results are qualitatively consistent with the chemical pressure effect [29]. Therefore, we note that the application of a magnetic field to "chemically pressurized" compounds with a combination of $E$-type Mn ordering and noncollinear rare-earth moments with polar magnetic point groups, such as orthorhombic HoMnO$_3$ [20], might enhance the ferroelectric polarization through polar to nonpolar phase transitions of the rare-earth spins.

In conclusion, we performed neutron diffraction experiments under high pressures and high magnetic fields using a single crystal of the multiferroic compound TbMnO$_3$. We found that the $E$-type ordering of Mn spins induces giant ferroelectric polarization over all the parameter regions investigated, through an exchange striction mechanism. We propose that the magnetic-field-induced polarization enhancement occurs via a mechanism in which the polarization, reduced by polar ordering of the Tb moments in a zero field, is recovered through the field-induced transition from polar to nonpolar Tb orderings at $H_{\|a} \sim 2$ T. Finally, we note that the application of magnetic fields to chemically pressurized manganites may also enhance ferroelectric polarization even at ambient pressure, via the polar to nonpolar transition of rare-earth orderings.

The authors wish to thank T. Aoyama and T. Kimura for helpful comments. The images shown in Fig. 1 were depicted using the VESTA [30] software program developed by K. Momma. This work was supported by JSPS KAKENHI Grant No. 15H05433 as well as the TUMOCS project, which has received funding from the European Union Horizon 2020 Research and Innovation Program under the Marie Sklodowska-Curie Grant Agreement No. 645660.